\documentclass[12pt]{article}

\usepackage{amssymb}

\usepackage{hyperref} 

\usepackage[autostyle=true,german=quotes]{csquotes} 

\usepackage{amsmath,amssymb,amsthm,amsfonts,amsbsy,latexsym}

\usepackage{caption}
\usepackage{subcaption}

\usepackage{longtable}
\usepackage{algorithmic}
\usepackage[algoruled,german]{algorithm2e}

\usepackage{graphicx}

\usepackage{wrapfig}
\usepackage{graphicx} 
\usepackage{tikz}
\usetikzlibrary{shapes.geometric, arrows}
\tikzstyle{startstop} = [rectangle, rounded corners, minimum width=3cm, minimum height=1cm,text centered, draw=black, fill=blue!30]
\tikzstyle{io} = [trapezium, trapezium left angle=70, trapezium right angle=110, minimum width=3cm, minimum height=1cm, text centered, draw=black, fill=blue!30]
\tikzstyle{arrow} = [thick,->,>=stealth]
\usepackage{tabulary}
\usepackage{tabularx}

\usepackage{multirow}
\usepackage{array}
\usepackage{hhline}
\usepackage{xcolor}
\usepackage{booktabs}
\usepackage{longtable}
\usepackage{siunitx} 
\usepackage{csvsimple}

\usepackage[normalem]{ulem}
\usepackage{listings}

\definecolor{codegreen}{rgb}{0,0.6,0}
\definecolor{codegray}{rgb}{0.5,0.5,0.5}
\definecolor{codepurple}{rgb}{0.58,0,0.82}
\definecolor{backcolour}{rgb}{0.95,0.95,0.92}

\lstdefinestyle{mystyle}{,   
    commentstyle=\color{codegreen},
    keywordstyle=\color{codegreen},
    numberstyle=\tiny\color{codegray},
    stringstyle=\color{codepurple},
    basicstyle=\ttfamily\scriptsize,
    breakatwhitespace=false,         
    breaklines=true,                 
    captionpos=b,                    
    keepspaces=true,                 
    numbers=left,                    
    numbersep=4pt,                  
    showspaces=false,                
    showstringspaces=false,
    showtabs=false,                  
    tabsize=2
}

\usepackage[nameinlink,capitalise,german]{cleveref}

\newcommand{\mytilde}{{\raise.17ex\hbox{$\scriptstyle\mathtt{\sim}$}}\xspace}

\usepackage{mathtools}

\newcommand*\samethanks[1][\value{footnote}]{\footnotemark[#1]}

\usepackage{color}

\usepackage[nottoc]{tocbibind}

\usepackage{natbib}
\usepackage{hyperref}                   
\hypersetup{                            
    colorlinks=true,                    
    linkcolor=black,                     
    citecolor=black,                     
    urlcolor=black                       
}

\title{Text Data Analysis and Classification Methods - Insights from Customer Letters in Life Insurance}
\author{Andreas Groll\thanks{Department of Statistics, TU Dortmund University}, Marie Punsmann\samethanks, Leonid Zeldin$^{*,}$\thanks{Corresponding author: leonid.zeldin@tu-dortmund.de; Department of Statistics, TU Dortmund University}}
\date {April 2026}

\begin{document}

\maketitle

\mbox{}
\vfill

\noindent Statements relating to the ethics and integrity policies: \\~\\

\noindent Unfortunately, our data are confidential and not available. However, if requested by the referees, we could make example code available, which illustrates the usage of our fitting procedures. Furthermore, we have no funding and no conflict of interest to declare.
\pagebreak

\begin{abstract}



\noindent 

\noindent The business of life insurance companies is characterized by long-term contracts. For this reason, data describing customers is of immense value. A portion of the data provided to the customer is rarely or not at all analyzed. This includes customer letters of any kind. This work focuses on classifying customer letters as cancellations and identifying the respective reason, if available. The outlined approach can also be applied to other business transactions and reasons. We discuss data acquisition and preparation, present alternatives, and explain the reasons for the chosen approach. A successful implementation of such a tool can lead to a better understanding of customer cancellation behavior by the insurer, enabling more targeted actions in certain situations.


\end{abstract}


\section{Introduction}

The longevity of life insurance contracts makes it imperative for an insurer to better understand its portfolio. This is achieved through numerous analyses. The more data that can be incorporated into such analyses, the better and more precise the statements about its customers, as explained in detail in \cite{Insurance_Customer_Touching_Points}. This automatically enables the insurer to calculate the risk and the policy premiums more accurately and adapt better to certain situations. It also allows for the recognition of certain indicators and early preventive actions. Moreover, the literature points even to the relevance of collaborations with external partners, which support access to non-insurance customer data to consequently
attract customers \citep{external_data}.

From these considerations, it emerges that every piece of information the insurer can obtain from its customers and is allowed to use holds enormous value. As extensively outlined in e.g. \cite{Insurance_2030} and \cite{Development_AI_financial_System}, the utilization of AI for data acquisition and processing is indispensable for insurance companies. A highly underestimated and largely unexplored field is the analysis of customer letters received by the insurer, see \cite{document_info_extraction}. Typically, these are read, addressed, and then archived. Occasionally, there may be manual or rule-based machine classification of the letters before archiving. As a result, a considerable amount of potentially high-value data provided by policyholders is buried in the archives. Some companies have already imlemented fraud detection tools, which use e.g. these letters but also other communication channels to defend the company against the fraud attacks \citep{AI_in_finance}. Furthermore the extraction of text information and the analysis of this data is already an usual process for documents or the chatbot communication, see, e.g. \cite{brief_survey} and \cite{AI_in_insurance}.

This work aims to take the first step in integrating these data into a process, extracting and analyzing them. \cite{extraction_old} already demonstrated the importance of this procedure. Due to the high level of burocracy and the lack of automation in the insurance sector it seems to be a very usefull application for the next years. Numerous customer letters are examined, using image and text recognition tools to extract information. These findings can facilitate customer segmentation, thereby enhancing the company's prospects for cross-selling and fostering effective communication \citep{comment_ai}. Several machine learning classification methods are then employed to categorize these letters, allowing them to more or less self-learn the rules for classification.

Next, the letters will be investigated regarding the  purpose for their creation. Specifically, cancellation letters will be considered, separated from other letters, and subsequently analyzed with respect to the reasons customers provide for their cancellations. A significant amount of work has already been dedicated to predicting customer churn behavior, see, e.g. \cite{churn_pred} and \cite{churn_pred_comparison}. This underscores the significance of this research area.

Once this information is extracted, it can be added to the databases. The same approaches can be extended to other customer letters and their corresponding reasons. This would result in the insurer optaining a better understanding of customer actions, leading to the aforementioned benefits. For example the insurer could offer personalized services \citep{Personalization} to increase the satisfaction of the customers or use the outputs to improve the identification of potential new business \citep{cross_selling}.

The remainder of the manuscript is structured as follows. In Section~\ref{sec:motiv}, the basic idea of this work is further elaborated. This is followed by a detailed description of document processing in Section~\ref{sec:data}, covering data acquisition, privacy concerns, technical requirements, and the final step of data preparation: labeling, which proves to be the most labor-intensive step. Section~\ref{Texte} explains how to extract texts from the documents, addressing methods, associated difficulties, and the approach taken with the available data. The specific procedures and the corresponding used methods are prestendet in Section~\ref{Methoden}, followed by their results in Section~\ref{Auswertung}. Finally, the work is summarized in Section~\ref{Zus_und_Ausblick}, and potential future steps are outlined.

\section{Motivation}\label{sec:motiv}

An insurance is meant to secure an insured party against the occurrence of an undesirable event by spreading the risk across the so-called collective, composed of all policyholders of a specific tariff. If this event occurs for an insured party, it is not the individual alone who pays, but the entire mentioned collective \citep{schmidt2002}.

Insurance companies must adhere to many regulations and laws to ensure that policyholders can secure their insurances under fair conditions, see, e.g. \cite{regulatory}. In Germany, the Federal Financial Supervisory Authority (German: BaFin) takes on the supervision of insurance. Among other responsibilities, it is accountable for protecting policyholders and beneficiaries of insurances, ensuring that insurance companies are always able to fulfill their obligations.

The present work deals with life insurance. A portion of life insurances are policies that provide coverage against biometric risks such as death or disability. Another portion includes pension or investment-linked life insurances. The peculiarity of this sector lies in the fact that insurance contracts often exist for many years and are rarely substantially altered, resulting in limited interactions between the insurance company and its customers. Therefore, it is crucial for the insurer to make the most of these rare contacts with policyholders to better understand the insurance portfolio. One underutilized form of contact occurs in written form, through which the company can gather additional information to complement existing contract and payment data. This can be achieved by understanding the reasons behind certain business events and incorporating them into deeper analyses.

For this purpose, customer letters from a major German life insurance company (name not provided for confidentiality reasons) regarding the following five different business events should be examined: cancellation, premium pause, premium waiver, increase, and policyholder change. Then, a list of possible reasons should be created for each business event, and the business events should be classified into the respective reasons. This differs significantly from the aforementioned paper \citep{document_info_extraction}, where the authors explored numerous methods to analyze various documents and extract key customer information, but did not delve deeper into customer communication.

Before requesting these data, an initial check was carried out to determine whether it is legally permissible to conduct the planned analyses with the data. In doing so, both the company's internal data protection officer and the responsible entity for the data at the life insurance company were contacted, and a thorough procedure was coordinated and planned.

Through the content analysis of the data, it has been noticed that the existing classification of business events is flawed, leading to adjustments in the questions related to data quality. As shown in \cite{classification_data}, the problem of document classification is very current in the world of data, and there are several methods to improve the manual or rule-based automatic processes. In the present work, we focus on the letters labeled as cancellation, and on the classification task for this particular category.

The final goal of this work can be divided into two parts. In the first part, a process will be developed that identifies document on cancellations from the entire set of documents. If possible, it should also output the likelihood of a cancellation. It cannot be ruled out that among the letters of other business events, cancellations may also be found, so this classification does not start with perfectly labeled data. However, due to the fact that we processed all letters in advance in an internal process and sorted into the respective category to the best of our knowledge and expertise, we believe that this problem can be neglected.

The second goal is to determine the reason for cancellation based on the cancellation letters. Probabilities of each cancellation reason will be provided, classifying them into the four most common ones\footnote{``change of profession", ``financial reasons", ``retirement" and ``death"} together with a fifth category denoted as ``other reason/no reason." The desired final output is illustrated in \autoref{bsptb}.
\begin{table}[!htp]
\centering
\begin{tabular}{cccc}
\toprule
Letter ID & Cancellation & Probability & Reasons \\ \midrule
Letter 1 & No & - & - \\ 
Letter 2 & Yes & - & Financial Reason \\ 
\hspace{0.01cm}\vdots &\hspace{0.01cm}\vdots & \hspace{0.01cm}\vdots & \hspace{0.01cm}\vdots \\ \bottomrule
\end{tabular}
\caption{Example rows of the output table}
\label{bsptb}
\end{table}

During the last few years, the growth of AI tools and the possibilities they bring have been immense. Alongside this development, an important challenge has emerged: the risk of discrimination and unfairness. Working with such sensitive data creates many opportunities for such issues to arise.

In this context, direct discrimination must be considered, as it can arise from personal attributes that may serve as straightforward proxies. To mitigate this risk, a dedicated role within the data science team should be established. This role would involve close interaction with the data and responsibility for monitoring each step of the development process, with the aim of protecting individuals from discriminatory outcomes. Measures may include comprehensive documentation as well as the use of multiple evaluation metrics to assess whether newly developed models exhibit discriminatory behavior or algorithmic bias.

In addition to the classification of data carried out in this work, sensitive stages of AI-driven processing should ideally be reviewed by domain experts to reduce the risk of critical errors that could harm reputation or lead to regulatory issues. Comprehensive studies such as \cite{bias_DAV} outline measures that can be implemented to mitigate discrimination and algorithmic bias. These measures can support and extend the model development process described in this work.

However, neither discrimination nor algorithmic bias is directly addressed in this study, as its sole objective is to extract specific information and classify it into predefined categories without assigning value judgments to the results.

Beyond the previously discussed risks, additional threats arise when analyzing free-form customer communications. It is important to note that, once customers become aware of research efforts aiming to improve processes based on their letters, they may intentionally manipulate the content, ranging from misleading to fraudulent statements. Detecting such manipulations is particularly challenging, given that such disclosures are voluntary and fully at the client's discretion.

Recent research provides valuable methodologies tailored to this context. For example, sentiment analysis of CEO letters to shareholders has been used to detect fraud by analyzing adjective usage patterns paired with SVM classification \citep{ceo_letter_sentiment}. Similarly, fine-grained element identification in fraud-related complaint texts employs a BERT-based model (BERT stands for Bidirectional Encoder Representations from Transformers) to classify individual clauses according to fraud-related roles \citep{complaint_text_fine_grained}. Furthermore, fraud investigators have leveraged deep-learning techniques to detect coded language in email communications, achieving high accuracy with BERT models \citep{codeword_detection}. While integrating these advanced approaches is technically feasible, doing so may exceed the scope of the present study, and potentially the broader research project, since the current work treats customer letters as indicators rather than definitive decision-making inputs.

This is primarily an exploratory work aiming to investigate whether modern classification approaches from statistical and machine learning are able to find cancellations and their reasons using the existing data. Therefore, the focus is on testing various methods for information retrieval rather than constructing a perfectly adapted model. \\
\\

\section{Data} \label{sec:data}

In order to be able to perform a thorough statistical analysis, suitable data must be acquired in advance, and all topics related to the data must be clarified. Since we analyze life insurance data, leading to many applications being highly regulated, the data protection topic must also be thoroughly addressed. In the following sections, we will focus on these aspects specifically.

\subsection*{Data Acquisition} \label{acquisition}

In line with typical practices in large corporations, obtaining data proved to be a convoluted and time-consuming process. After requesting the data from the responsible data unit, it took several
months until actual access to the data could be obtained. Furthermore, the quality of the data did not meet all expectations, leading to an adjustment of the original goal of the project. This is explained in more detail in Section~\ref{DuDLAB}.

The insurer provided us with 14,575 customer letters, which approximately corresponds to the number of letters received over the past five years. In \autoref{overview} in the appendix, the number of documents for each business transaction is presented. It is evident that there are the fewest documents for contribution exemption ($641$), while most documents refer to policyholder changes ($6,579$). However, since these numbers were found to be inaccurate during the course of investigation, the exact figures are not examined in detail here. The documents consist of scanned or otherwise processed files, presented as protected, non-searchable PDFs. Therefore, the text from these files must be initially extracted for processing. The documents vary in length, ranging from single-page letters to documents exceeding $40$ pages.

\subsection*{Data Protection} \label{Datenschutz}

For the conclusion of contracts, calculation of contract conditions, and administration of contracts, insurance companies must use personal data. This also includes sensitive data, such as health information, which needs to be evaluated when a life insurance policy is taken out. Protecting this data is crucial. To ensure that insured individuals can rely on this protection, the German Insurance Association has largely established uniform standards for data protection in its Code of Conduct (CoC), see \citet{CoC}.

A significant number of German insurance companies, including the represented life insurer here, have committed to adhering to this Code of Conduct. All rules outlined in this code have been established in compliance with the General Data Protection Regulation (GDPR), the Federal Data Protection Act (German: BDSG), see \citet{BDSG}, and all relevant sector-specific regulations on data protection. Additionally, principles of transparency, necessity of processed data, and data minimization are particularly emphasized.

Currently, reasons for business transactions are not yet recorded at the represented life insurer. However, these are important, as, for example, lapse rates are regularly estimated as part of actuarial calculations. If, however, the lapse rates were higher due to the pandemic in 2020, this should be considered in forecasts for future years. Therefore, analyses from an actuarial perspective provide valuable insights and are allowed to be conducted under Article 2 of the CoC.

Based on Article 47(5) of the GDPR and Article 3(2) of the CoC, this work pays particular attention to not storing personal data. Consequently, graphical representations of analyses based on the data, such as word clouds, are not possible.

Storing the data in a dedicated folder accessible only to the employees directly involved with the present use case aligns with Article 4 of the CoC. Finally, the request for customer letters related to the specified business transactions and their further processing are carried out in coordination with the Data Protection Officer, as regulated in Article 27 of the CoC.

\subsection*{Technical Requirements}

In addition to complying with data protection regulations, another crucial prerequisite for the work is that certain technical requirements have to be met in order to construct the process. For this purpose, \texttt{Python} was chosen as the programming language. The overview of the used packages can be found in Table \ref{Paketübersicht} in the appendix. 

The installation of \texttt{Python} packages was sometimes either very complicated or not possible, as at that time, the insurance industry was quite skeptical about open-source solutions, and all extensions had to undergo complex internal approval processes. Consequently, some packages had to be foregone, and work had to proceed with alternative packages.

\subsection*{Labeling} \label{DuDLAB}
To use the given data for training and testing in a supervised classification task, each observation must have a known true class. In this study, business transactions are already archived by the life insurer, but no information is available regarding the content of the letters. Therefore, cancellation letters must be individually examined and manually labeled to identify the cancellation reasons. This work focuses exclusively on cancellations, as they are particularly relevant for forecasting lapse rates. Of the $3,477$ cancellations identified, insurance numbers, risk numbers, and cancellation reasons—if available—are recorded. Insurance and risk numbers uniquely identify contracts and are crucial for merging with basic data later.

During preprocessing, the classification of business transactions is re-checked, and incorrectly classified documents are addressed. Some documents belong to other transactions, such as premium exemptions, customer inquiries, or business terms, and are reassigned to maintain dataset size. Internal emails or attachments unrelated to the actual cancellation are excluded. These documents, often repetitive or irrelevant, are stored separately and not used further. Duplicated documents are also removed by performing a full comparison of the imported data, reducing multiple occurrences to a single copy, as detailed in the beginning of Section~\ref{Auswertung}. For cancellations, additional checks ensure duplicates based on insurance and risk numbers are eliminated. However, cancellations with identical insurance numbers but without risk numbers are temporarily retained, as supplementary contracts might also be affected. This ensures minimal redundancy in the training and test datasets.

For cancellations with a reason provided, reasons are initially labeled and later grouped into broader categories. Categories with fewer than 10 occurrences are grouped under ``other reasons", which includes cases such as cancellations due to the pandemic or relocation abroad. These are treated as missing reasons and merged into the category ``other reason/no reason" for analysis. Section~\ref{kg} explains why there is no differentiation between the categories ``other reasons" and ``no reason" in this context. An overview of these categories is presented in \autoref{Kgründe}. Although cancellations can theoretically have multiple reasons, in this dataset, only one primary reason is assigned per cancellation. Secondary reasons categorized as ``other reasons" are excluded, as they cannot be identified in the analysis.

After processing, $12,092$ documents remain, including $1,257$ cancellations, of which only $83$ have one of the considered cancellation reasons. This imbalance is addressed using an oversampling method for most analyses, except for the word search. Most cancellations lack a stated reason, as there is no obligation to justify a cancellation. This uneven class distribution complicates classification and necessitates oversampling to balance the dataset. The presented numbers (including those in \autoref{Kgründe}) reflect the counts after all duplicates and irrelevant documents were removed.

\begin{table}
\centering
\begin{tabular}{lr}
\toprule
Reason for Cancellation & Frequency \\ \midrule
Change of Profession & $23$ \\
Financial Reasons & $26$ \\
Retirement & $18$ \\
Death & $16$ \\
Other Reason/No Reason & $1,174$ \\  
\bottomrule
\end{tabular}
\caption{Absolute frequencies of the different observed cancellation reasons}
\label{Kgründe}
\end{table}

\section{Extracting Text} \label{Texte}

The goal of this section is to explain how the texts from non-searchable PDF files are extracted and to describe the procedure for the given data. 
For this purpose, the PDF files must first be converted from protected to unprotected files \citep{protectunprotect}, as discussed above and clarified with regards to data protection. The next step is to extract the text from these unprotected PDF files using text recognition. However, \texttt{Python} cannot directly read these PDF files, which is why they are converted into image files in the PIL format \citep{PILFormat}. The analysis in this work is performed using \texttt{Python} version 3.9.7 \citep{pythonversion}. As an integrated development environment, \texttt{Spyder} version 5.1.5 from \texttt{Anaconda 3} is used. A complete list of all used packages\footnote{In this work, any combination of functions is referred to as a package. This includes modules and libraries.} can be found in the appendix.

\subsection{Text Recognition}

The final and most computationally intensive step of the procedure is the extraction of text from images using Optical Character Recognition (OCR). OCR aims to convert text in images into a machine-readable format for further processing.

For this task, the open-source tool \texttt{Tesseract} \citep{tesseract}, accessed via its \texttt{Python} implementation \texttt{pytesseract} \citep{pytesseract}, is used. This choice was guided by the tool’s availability, compatibility with \texttt{Python}, and its proven performance in related studies \citep{qualitytool1, qualitytool2}. \texttt{Tesseract} processes binary images (black-and-white) as input and performs connected component analysis to identify and classify text regions, including handling inverse text (white on black). 

A key feature of \texttt{Tesseract} is its ability to identify text baselines, enabling recognition of curved or irregularly aligned lines, as often seen in scanned documents. Text is divided into lines and further into words. For text without fixed spacing, such as handwritten or irregularly printed text, \texttt{Tesseract} uses word gap analysis and adaptive segmentation techniques to address ambiguities.

The classification process involves multiple stages. First, each word is compared to categories such as dictionary words, numerical values, and case-specific patterns (e.g., all upper case or title case). The word with the best match is selected based on minimal total distance. A static character classifier handles initial recognition, using a compact training dataset that generalizes well, even for incomplete or broken characters. Recognized words from the first pass are then used to train an adaptive classifier, which performs a second pass to improve recognition of previously unrecognized words \citep{tesseract}. This multi-step process ensures accurate text extraction, even for challenging inputs like irregularly spaced or partially degraded text.

\subsection{Application to the Present Data}

After applying this procedure to the available data, the data is presented in the form of a list. For single-page documents, the text of the respective letter is concealed behind the list elements, and for multi-page documents, another list is formed, with each element containing the text of a page.
For the subsequent processing, the pages are concatenated, and there is no longer a record of how many pages a letter has or where page breaks occur.
At this point, the multiply occurring documents are present in the used data, as text recognition is necessary for identifying duplicated letters.

Machine-written texts are read much more effectively by \texttt{Tesseract} than handwritten texts. However, during labeling and examination of the texts, it has been noticed that there are not many handwritten texts, and many handwriting styles, in general, are very difficult or even impossible to decipher. Therefore, the quality of the extracted data is considered acceptable and is not further optimized.

\section{Methods in Text Data Analysis and Classification} \label{Methoden}

Preprocessing of the texts is necessary for further analysis. In this process, the key words and phrases, along with their frequencies across all texts, are stored. Subsequently, classification is performed using various models. The methods required for this purpose are introduced in this section.

\subsection{Text Data Analysis} \label{Dataanalysis}
At the beginning of this subsection, some terminology is introduced, which will be used in this work. The set of all $m$ texts to be examined, representing individual writings in this context, is referred to as the \textit{corpus}. For content analysis, the texts need to be divided into words, sentences, or word groups. In this work, \textit{tokenization} on a word basis is used for this purpose. In this process, a vector consisting of \textit{tokens} is created from a coherent text. These tokens can be single or multiple consecutive words as well as sentences citep{token}. In the case of $n$ consecutive words, this token is also called an \textit{$n$-gram}. A 2-gram is also known as a \textit{bigram}, and a 3-gram as a \textit{trigram}.

The so-called \textit{stemming} achieves a reduction in the number of tokens by breaking down or reducing words to their word stems \citep{stemming}. This has the advantage of grouping together words that share the same word stem. Additionally, \textit{stop words} can be removed before the analysis as they are irrelevant \citep{stopwords}. 

\begin{table}
\centering
\begin{tabular}{lllll}
\hline
&Token 1& Token 2 & \ldots & Token $n$ \\ \hline
Document 1 &$h_{11}$& $h_{12}$ & \ldots &$ h_{1n} $\\
Document 2 &$h_{21}$& $h_{22}$ & \ldots &$ h_{2n} $\\
\hspace{0.01cm}\vdots & \hspace{0.01cm}\vdots & \hspace{0.01cm}\vdots & $\ddots$ & \hspace{0.01cm}\vdots \\
Document $m$ &$h_{m1}$& $h_{m2}$ & \ldots & $h_{mn}$\\ \bottomrule
\end{tabular}
\caption{General structure of document-term matrix}
\label{dtmbsp}
\end{table}

Another important tool for processing text data is the \textit{Document-Term Matrix}.
The structure of a Document-Term Matrix is illustrated in \autoref{dtmbsp}, where the frequencies $h_{j \omega}$ of the occurrence of token $\omega$ in the $j$-th document are summarized for all considered documents and tokens, regardless of their order \citep{gini}.

\subsection{Feature Selection}

In the context of feature selection, \citet{gini} introduce the Gini index as a measure to evaluate the discriminatory power of tokens after stop words have been removed. The Gini index assesses how well a token distinguishes between different classes, with higher values indicating stronger association with a specific class. It accounts for the distribution of a token across classes, identifying tokens that occur disproportionately in one class compared to others. For datasets with imbalanced class sizes, a normalized version of the Gini index can be used to ensure fair comparison. This approach helps prioritize tokens that are most relevant for classification tasks.

\subsection{Classification Approaches}

In this subsection, the two classification methods used in this work, Random Forests and Support Vector Machines, are introduced. These methods are generally applicable and do not specifically relate to text data. For classification, the data is in the form of a Document-Term Matrix.

\subsubsection*{Word Search} \label{Word_Search}

The Word Search methodology is a simple and interpretable approach for text classification that relies on matching predefined tokens indicative of specific categories. The process involves defining lists of tokens strongly associated with each category and then classifying documents based on the presence or absence of these tokens.

This methodology is particularly useful in scenarios where interpretability and computational efficiency are priorities. Unlike more complex machine learning models, Word Search relies on manually selected tokens, making the classification process straightforward and understandable. Tokens are chosen based on exploratory analysis of the training data, such as observing token frequencies or creating word clouds. These selected tokens serve as clear indicators of the categories under consideration.

While Word Search is computationally efficient and provides clear results, it has limitations. The method depends heavily on predefined tokens, making it less adaptable to complex datasets or cases where context is necessary to determine meaning. It also requires domain knowledge for effective token selection.

This approach aligns with fundamental principles of rule-based classification in text mining. For instance, Boolean retrieval, as discussed in \cite{manning2008introduction}, forms the basis for matching documents containing specific terms. Additionally, \citet{sebastiani2002} and \citet{sarawagi2008} have described similar rule-based methods in the contexts of text categorization and information extraction, highlighting the utility of keyword-driven techniques for interpretability and efficiency.

\subsubsection*{Random Forest} \label{Random Forest}

Random Forests are ensembles of decision trees used for classification tasks, including text recognition. In this work, each tree is built using a binary splitting process to minimize impurity (measured by metrics like the Gini index, see \citealp{Breiman_classification_trees} and \citealp{elements_stat_learning}). Trees are grown on random bootstrap samples from the training data, and splits at each node are determined using a random subset of features to introduce variability and reduce correlation between trees (see \citealp{breiman_random}). The final prediction is made by aggregating the outputs of all trees through majority voting, a technique called bagging \citep{breiman_bagging}. This approach reduces variance and improves the robustness of predictions compared to a single decision tree.

\subsubsection*{Support Vector Machines}
Support Vector Machines (SVMs) are a classification method based on finding an optimal hyperplane to separate data into distinct classes, see \cite{svm}. For linearly separable data, SVMs identify the hyperplane that maximizes the margin between classes, with support vectors defining the boundary. When data is not perfectly separable, SVMs introduce soft margins using slack variables to allow some misclassifications while minimizing overall error. To handle non-linear relationships, SVMs use kernel functions, which map the data into higher-dimensional spaces where it becomes linearly separable. A commonly used kernel is the Radial Basis Function (RBF; e.g., the Gaussian kernel, \citealp{radial_kernel}), which is effective for capturing complex patterns in data without requiring prior knowledge of specific relationships.

\subsubsection*{Oversampling}

When dealing with data sets containing uneven class sizes, it is useful to employ oversampling. In this work, random resampling is used. This involves enlarging the smaller class by drawing with replacement from the original data set until the desired sample size is reached (for details, see e.g., \citealp{LearningfromImbData,he2009learningfromimbdata}).

\subsubsection*{Youden Index}

Given the imbalance in the data in this study, the Youden Index is employed to assess the classification performance and determine the optimal threshold for the model. The Youden Index is a widely used metric for evaluating classification performance, especially in imbalanced datasets (see \citealp{youden}). It is defined as the sum of sensitivity (the proportion of correctly classified instances within one class) and specificity (the proportion of correctly classified instances within the other class) minus one. The index ranges from $-1$ to $1$, with higher values indicating superior performance. Additionally, the Youden Index is instrumental in identifying the optimal threshold by achieving a balance between sensitivity and specificity. Once the threshold is established, the results can be summarized in a classification table, as illustrated in \autoref{ktab}.

\begin{table}[!htp]
\centering
\begin{tabular}{lccc}
  \toprule
  & \multicolumn{2}{c}{Prediction} & \\\cmidrule{2-3} Group & Class Zero & Class One &   $\sum$ \\
 \midrule
Class Zero & $m_{00}$ & $m_{01}$ & $m_{0.}$   \\ 
Class One & $m_{10}$ & $m_{11}$ & $m_{1.}$ \\ 
$\sum$ & $m_{.0}$ & $m_{.1}$& $m$   \\ 
   \bottomrule
\end{tabular}
\caption{General structure of classification table}
\label{ktab}
\end{table}

\section{Results} \label{Auswertung}

After the data preparation is complete, the further analysis is divided into two parts. The first part involves classifying the documents into ``Cancellation" and ``No Cancellation." The second part focuses on assigning the cancellation reasons on the subset of those documents which have been classified as ``Cancellation" in the first part of the analysis. Therefore, in this part, only the cancellations are considered.

\subsection{Locating Cancellations} \label{knk}

In the first step, the texts are prepared using methods outlined in Section~\ref{Dataanalysis} for the classification models subsequently applied. Following this, a word search, a Random Forest, and a Support Vector Machine are tailored to the data to classify the texts into the categories ``Cancellation" and ``No Cancellation."

\subsubsection{Text Data Preparation} \label{tm}

The first steps in text processing involve removing punctuation and converting all uppercase letters to lowercase. The texts are then tokenized on a word basis. Stemming is applied to reduce words to their root forms, followed by the removal of stop words. All numbers are also removed; if a number is part of a word, only the number is deleted, retaining the word itself.

To prepare the data for analysis, the dataset is divided into $70\%$ training data and $30\%$ test data. This division ensures a consistent proportion of cancellations and other documents in both sets, see Section~\ref{kg}. Using the same split for all research questions avoids overlap between training and test data at different stages of the analysis.

After initial processing, the corpus still contains a large number of tokens. To manage this, only words occurring in at least one percent of cancellations and other documents are retained. This step removes most names and helps comply with privacy regulations by excluding unnecessary personal data, see Section~\ref{Datenschutz}.

To reduce the number of tokens further, Document-Term Matrices (DTMs) of monograms and bigrams are created, as these are sufficient for the analysis. Higher $n$-grams are not used to limit computational complexity. Tokens are filtered based on their Gini index; only those with an index less than one are retained, which excludes tokens that appear exclusively in one class. Although alternative methods for addressing incorrectly recognized words, described in Section~\ref{othermodels}, exist, they are not utilized in this study.

A joint DTM of mono- and bigrams is constructed for the training data and serves as the final matrix for the analysis. For the test data, a corresponding DTM is created using only the tokens selected from the training data.

\subsubsection{Classification with Word Search} \label{wordsearch}

The Word Search method, presented in Section~\ref{Word_Search}, starts with token selection from the training data, using word clouds created for both cancellations and non-cancellations, and does not require the data preparation steps outlined in Section~\ref{tm}. These word clouds visually highlight frequently occurring tokens, though many are not specific to cancellations or business transactions. To ensure meaningful token selection, only those that are semantically relevant to cancellations are retained. However, since the word clouds include sensitive information, a visual representation cannot be provided\footnote{Due to confidentiality, the word clouds are not shown in this work.}.

At the beginning, the most prominent tokens semantically fitting a cancellation are manually selected. However, choosing the right tokens is a challenge because some terms appear in both cancellations and non-cancellations. Using word clouds alone is not enough to select the most meaningful tokens for classification. After considering the frequency of tokens in the training data, two lists are created: one for cancellation-specific tokens and one for non-cancellation tokens.

The process of token selection follows the principle of Occam's razor\footnote{The simplest explanation for a phenomenon is preferred over more complicated ones.}, avoiding unnecessary complexity and focusing only on the most relevant tokens. In total, 12 tokens are chosen: 8 monograms and 4 bigrams, all based on their frequency in the cancellation class\footnote{Again, a visual representation is omitted, as these are German word stems that do not make sense in English.}. Only tokens that occur more frequently in cancellations than in non-cancellations are retained.

The classification performance is evaluated using the selected tokens. In the training phase, the model achieved a sensitivity of $0.927$ and a specificity of $0.942$, resulting in a Youden Index of $0.869$, as shown in Table~\ref{KTesttab}. This performance improves when additional tokens specific to non-cancellations are included, raising the specificity to $0.956$ and increasing the Youden Index to $0.884$ (see Table \ref{WSTraintab}).

Further refinement of the model using these tokens leads to an increase in specificity, reaching $0.965$ with a Youden Index of $0.890$. The final model is tested on unseen data, as shown in Table \ref{WSTesttab}, and achieves a sensitivity of $0.920$ and a specificity of $0.966$, demonstrating consistent performance across both training and test sets.

\begin{table}[h]
\begin{tabular}{cccc}
\toprule
& \multicolumn{2}{c}{Predicted Class}     &      \\ \cmidrule{2-3}
True Class & Cancellation & No Cancellation & $\sum$ \\ \midrule
Cancellation & $816$ & $64$ & $880$ \\
No Cancellation & $441$ & $7,143$ & $7,584$ \\
$\sum$ & $1,257$ & $7,207$ & $8,464$ \\ \bottomrule
\end{tabular}
\centering
\caption[Result of the classification of the Word Search, which only looks for cancellation words (excluding the word stem of ``cancel" in past tense) on the training data]{Absolute classification results of the Word Search, which only looks for cancellation words (excluding the word stem of ``cancel" in past tense) on the training data with sensitivity$=0.927$ and specificity$=0.942$}
\label{KTesttab}
\end{table}

\begin{table}[h]
\begin{tabular}{cccc}
\toprule
& \multicolumn{2}{c}{Predicted Class}     &      \\ \cmidrule{2-3}
True Class & Cancellation & No Cancellation & $\sum$ \\ \midrule
Cancellation & $814$ & $66$ & $880$ \\
No Cancellation & $267$ & $7,317$ & $7,584$ \\
$\sum$ & $1,081$ & $7,383$ & $8,464$ \\ \bottomrule
\end{tabular}
\centering
\caption[Classification of the Word Search, which looks for cancellation words and additional five tokens (per no cancellation), on the training data]{Absolute classification results of the Word Search, which looks for cancellation words and additional five tokens (per no cancellation), on the training data with sensitivity$=0.925$ and specificity$=0.965$}
\label{WSTraintab}
\end{table}

\begin{table}[h]
\begin{tabular}{cccc}
\toprule
& \multicolumn{2}{c}{Predicted Class}     &      \\ \cmidrule{2-3}
True Class & Cancellation & No Cancellation & $\sum$ \\ \midrule
Cancellation & $347$ & $30$ & $377$ \\
No Cancellation & $110$ & $3,141$ & $3,251$ \\
$\sum$ & $457$ & $3,171$ & $3,628$ \\ \bottomrule
\end{tabular}
\centering
\caption[Classification of the final Word Search on the test data]{Absolute classification results of the final Word Search on the test data with sensitivity$=0.920$ and specificity$=0.966$}
\label{WSTesttab}
\end{table}

\subsubsection{Classification with Random Forest}

In the Random Forest classification, oversampling is applied to balance the training data, with the cancellation group being half the size of the other documents\footnote{The following classifications are mainly performed using the \texttt{scikit-learn} package in \texttt{Python}. The \texttt{imbalanced-learn} package is also used for oversampling.}. The classification results are evaluated using sensitivity, specificity, and the Youden index, ensuring meaningful outcomes even with this adjustment.

The models are trained using default parameters in \texttt{Python} ($B = 100$, $n_{min} = 1$, and $mtry = \sqrt{p}$). The so-called ``large" model refers to the initial model created using the full Document-Term matrix from Section~\ref{tm}, which includes all available tokens from the corpus. However, many of these tokens are irrelevant to the classification task, resulting in a model that struggles to properly distinguish between cancellations and non-cancellations. In this model, sensitivity is $0.496$, and specificity is $0.503$. This distribution is reflected in \autoref{RF2Testtab} in the appendix.

To improve upon this, a smaller Document-Term matrix is considered, where only the tokens selected in the Word Search methodology (as described in Section~\ref{wordsearch}) are included. This model, referred to as the small model, benefits from two key advantages: it is faster to fit, and it includes only tokens that are relevant to distinguishing cancellations from non-cancellations. However, as discussed in Section~\ref{wordsearch}, token selection was somewhat subjective and may have missed other useful tokens that could further improve the model.

In terms of performance, the small model demonstrates better results. For the training data, the sensitivity is $0.924$, specificity is $0.973$, and the Youden index is $0.897$, showing a notable improvement compared to the large model. These results can be found in \autoref{RFTraintab} in the appendix. 

When evaluated on the test data, the small model again outperforms the large model. Sensitivity is $0.912$, specificity is $0.976$, and the Youden index is $0.889$, indicating more accurate classifications than the large model. The corresponding cross-table for the test data is presented in \autoref{RFTesttab} in the appendix. 

Therefore, based on both training and test data, the smaller Random Forest model proves to be the better model, likely due to its focus on more relevant tokens identified through the Word Search process.

\subsubsection{Classification with Support Vector Machine}
Finally, a classification with Support Vector Machines (SVMs) is performed\footnote{Default settings of \texttt{Python} and a radial basis kernel are used here.}. Similar to the Random Forests, a model is established on the Document-Term Matrix prepared in Section~\ref{tm} and one on the matrix with the selected tokens from the Word Search.

In \autoref{SVM2Traintab} in the appendix, the result of the classification with the large model on the training data is summarized, which looks exactly like the Random Forest result.
However, the classification of the test data, shown in \autoref{SVM2Testtab} in the appendix, is different. Here, documents are not randomly sorted into both classes, but a large portion of the documents in both classes is assigned to the ``no cancellation" class. This is also not a good classification. The sensitivity is very low at $0.143$, while the specificity at $0.863$ is within an acceptable range.
However, since a naive estimator that classifies all documents as ``no cancellation" would have a specificity of $1$, high specificity alone is not a measure of a good model.

In \autoref{SVMTraintab} in the appendix, the results of the classification in the small model with Support Vector Machine on the test data are depicted. Here, a similar pattern emerges as with the small Random Forest model, although some more documents are misclassified. The Youden index here yields $0.877$.

For the test data, whose results are found in \autoref{SVMTesttab}, the picture is somewhat different compared to the Random Forest: more cancellations are correctly classified, and more other documents are misclassified. Overall, a sensitivity of $0.926$ and a specificity of $0.968$ are obtained here. The Youden index of $0.894$ is just below $0.9$ and is thus good.

A comparison of the two SVM models shows that the small model delivers significantly better results here as well. Therefore, this will be compared with the small Random Forest model as well as the Word Search from Section~\ref{wordsearch}.

\subsubsection{Summary of Classification Results}

In \autoref{kfindvgl}, the sensitivities, specificities, and Youden indices of these three models after classifying the test data are listed.
It is evident that the Youden index of the SVM is the largest with a value of $0.894$. The sensitivity of the SVM is also the largest, with only the Random Forest achieving the best result in specificity. The Youden indices of the three models differ by a maximum of $0.07$, so no systematic difference can be detected here.

Due to the low number of features in all three models, they are not computationally intensive.
A clear choice for the best model cannot be made here, as all models are approximately equally suitable.

\begin{table}[h!]
\begin{tabular}{lrrr}
\toprule
Model & Sensitivity & Specificity & Youden Index \\ \midrule
Word Search & $0.920$  & $0.966$ & $0.887$\\ 
Random Forest & $0.912$ & $\textbf{0.976}$ & $0.889$\\
Support Vector Machine & $\textbf{0.926}$& $0.968$& $\textbf{0.894}$\\ \bottomrule
\end{tabular}
\centering
\caption{Overview of the classification results for all documents for the Word Search model and the respective best Random Forest and SVM models on the test data}
\label{kfindvgl}
\end{table}

\subsection{Cancellation Reasons} \label{kg}
To assign possible reasons to the cancellations, only the documents declared as cancellations are considered in the following part of the analysis. The four most common cancellation reasons, which occur most frequently and are classified here, are ``change of profession," ``financial reasons," ``retirement," and ``death." Additionally, there are many cancellations where no reason or another reason is given. It does not provide any added value to examine the ``other reason/no reason" class separately, as it does not reveal anything about the actual reason. Consequently, these cancellations, along with those without a specified reason, collectively form a fifth class.

\subsubsection{Text Data Preparation}
The first part of the text data analysis of cancellations follows a similar process to that of all documents (see Section~\ref{tm}). To investigate this question, cancellations are divided into a training and a test data set, comprising $70\%$ and $30\%$ of the cancellations, respectively. Just like in Section~\ref{tm}, the data sets are split in a way that ensures the cancellation reasons are represented in these proportions in both the training and test data sets.

The least common cancellation reason, ``death," occurs eleven times in the training data set and five times in the test data set with this split. Such a low number of documents must be handled with great caution. For instance, cancellations with this reason should not be easily removed from the data set. In contrast to text mining of all documents, cancellations constitute the total set of all documents when analyzing the reasons. Additionally, there are five classes instead of two. For example, when removing irrelevant tokens, a token is only excluded if it appears in less than one percent of all cancellations for each cancellation reason and in less than one percent of cancellations in the ``other reason/no reason" class.

At the end of this section, a Document-Term Matrix is created for the training data with selected tokens, and a corresponding matrix is generated for the test data, as in Section~\ref{tm}. The Gini coefficient is calculated for all classes. When computing the Gini index, many tokens again have a Gini index of one. Since there are more tokens than those considered in this part of the analysis, the subsequent models will not use these Document-Term Matrices but only the tokens selected in the Word Search (see Section~\ref{classreasons}).

\subsubsection{Classification} \label{classreasons}

The classification of reasons differs from the classification into ``cancellations" and ``non-cancellations" in that there are now five classes instead of two. This challenge is addressed by creating a separate binary model for each of the four reasons. Each model outputs the likelihood that the cancellation corresponds to the given reason. This way, each document is assessed to determine how well it fits into one of the four classes. If the probability of belonging to any of these four classes is too low, the document is categorized into the ``other reason/no reason" class through exclusion. This procedure is also known as ``one versus rest" classification. Finally, it is stipulated that there must be one procedure for all reasons, either a Random Forest, a Support Vector Machine, or a Word Search.

This section details all processing steps for the most common cancellation reason, ``financial reasons." All other reasons are processed using the same approach, with essential results presented in the appendix. A brief overview of the models for all cancellation reasons is provided at the end of the section.

For financial cancellation reasons, a manual Word Search is performed, similar to Section~\ref{wordsearch}, to create a list of relevant tokens. Six tokens are selected\footnote{Again, a visual representation is not meaningful here, as these are German terms reduced to their word stems and cannot be translated sensibly.} that most accurately classify the reason. If multiple token combinations apply, the one that correctly classifies the most cancellations is chosen. If there are still multiple combinations, the one with the lowest number of variables is selected. If multiple models persist, the model with alphabetically the earliest signal tokens is chosen.  Given the construction of four separate binary models, it is not meaningful here to search for tokens that do not speak for a specific reason, as already explained in Section~\ref{wordsearch}. Firstly, there may be documents that contain multiple cancellation reasons. Secondly, it is even more challenging here to manually find tokens that speak against a reason.

In \autoref{FWSTraintab} in the appendix, the classification table for the Word Search on the training data for financial reasons is presented. The sensitivity here is $0.722$, while the specificity is $0.988$. The classification on the test data is significantly worse with the Word Search, as shown in \autoref{FWSTesttab} in the appendix. The model only identifies two out of eight cancellations that originated from financial reasons, resulting in a sensitivity of $0.25$. The specificity is $0.992$, as expected due to low sensitivity.

The significant difference may be due to the fact that each document in the training data was examined, creating a form of unconscious overfitting to the training data. Furthermore, there are very few letters with this cancellation reason, so the model did not have enough data to learn potentially different formulations and later recognize them.

One option for oversampling cancellations is to use the same data set for all four models. This aligns with the aforementioned constraints, which is why this approach is taken here. Alternatively, a separate data set could be designed for each reason, with the class for the reason being made as large as for ``other reason/no reason".

For the following section, oversampling with a factor of four was performed for each reason. It should be specifically noted that this means the group without the respective cancellation reason has more training data in the Random Forest and the Support Vector Machine than in the Word Search.


When examining the Random Forest, it is noticeable that, as shown in \autoref{FRFTraintab} in the appendix, it classifies the training data well. With a sensitivity of $0.806$ and a specificity of $0.955$, a Youden index of $0.761$ is obtained. However, when looking at the test data, the results in \autoref{FRFTesttab} in the appendix indicate that the classification is significantly worse. Similar to the Word Search, the sensitivity here is only $0.25$. Along with a specificity of $0.970$, the Youden index is $0.220$.


The results of the classification with the Support Vector Machine on the training data are presented in \autoref{FSVMTraintab} in the appendix. Here, a sensitivity of $0.722$ and a specificity of $0.994$ are obtained. When applying the model to the test data, it is noticeable that only one cancellation is assigned to the financial reason (see \autoref{FSVMTesttab} in the appendix). This suggests that the weighting of sensitivity in the model for the objective is too weak.

\begin{table}
\begin{tabular}{lrrr}
\toprule
Model & Sensitivity & Specificity & Youden Index \\ \midrule
Word Search & $\textbf{0.250}$ & $0.992$& $\textbf{0.242}$\\
Random Forest & $\textbf{0.250}$ &$0.970$ & $0.220$\\
Support Vector Machine &$0.125$ & $\textbf{1.000}$&$0.125$ \\ \bottomrule
\end{tabular}
\centering
\caption{Overview of the classification results for cancellations into financial reasons or non-financial reasons for the Word Search model, Random Forest, and SVM}
\label{Fgfindvgl}
\end{table}

A comparison of the best models from Word Search, Random Forest, and SVM (see \autoref{Fgfindvgl}) reveals that the Word Search, based on the Youden index, is the best model. However, since all Youden indices are very low, it must be noted that no model classifies particularly well at this point.

Tables \ref{BALL}, \ref{RALL}, and \ref{TALL} from the appendix provide an overview of the classification results for other cancellation reasons. In \autoref{BALL}, it is noticeable that the classifications for job change test data are all very poor, and only the Word Search correctly assigns a cancellation to this reason.

\begin{table}
\begin{tabular}{lrrr}
\toprule
Model & Word Search & Random Forest & Support Vector Machine \\ \midrule
Financial Reason & \textbf{0.242} & 0.220 & 0.125 \\
Job Change & \textbf{0.100} & -0.014 & -0.014 \\
Retirement & 0.258 & 0.395 & \textbf{0.400} \\
Death & \textbf{0.597} & \textbf{0.597} & \textbf{0.597} \\ \bottomrule
\end{tabular}
\centering
\caption{Youden indices of the three models for all four cancellation reasons}
\label{Gfindvgl}
\end{table}

When comparing the Youden index for the test data of each model for all four cancellation reasons (see \autoref{Gfindvgl}), it is noticeable that the Word Search offers the best classification for both financial reasons and job change. For retirement, the Support Vector Machine is the best, closely followed by the Random Forest, while the Word Search is the least effective. For death, all models result in the same classification. Overall, it can be concluded that the classification for all four cancellation reasons is poor with all models.

Since the best classification is achieved by Word Search in three out of four cases, Word Search is selected as the final model for this step. In a random examination of cancellation reasons not detected by Word Search, it is noticeable that some of these documents are not well-read, and in some cases, no words were recognized. This leads to the documents not being assigned to a cancellation reason.

\section{Outlook} \label{Zus_und_Ausblick}

In the present study, an attempt was made to classify customer letters from a major German life insurance company and analyze the associated reasons for customer actions. Specifically, efforts were made to distinguish cancellation letters from other correspondence and examine four cancellation reasons: ``Change of profession," ``Financial reasons," ``Retirement," ``Death," as well as the category ``Other reason/No reason."

To facilitate this, the aforementioned life insurance company provided $14,575$ letters (see \autoref{overview}), consisting of scanned letters and archived emails in PDF format. Apart from addressing data privacy concerns, which needed to be clarified and strictly adhered to, and overcoming technical challenges within a large corporation with robust bureaucratic processes, extracting text from letters in PDF format proved to be a real challenge.

Once the letters were converted into text using available methods in \texttt{Python}, the preparation for their analysis began. The preparation involved a thorough comparison of letters and matching those with the same insurance number to eliminate duplicates or redundancies. Additionally, the texts were segmented into individual words (monograms) and word pairs (bigrams). These were then reduced to their word stems through stemming to eliminate variations in formulations. Finally, a set of stop words that generally do not contribute value to such analyses were removed.

For the classification of letters into ``Cancellation" and ``Non-cancellation," a more in-depth preparation was necessary. Only essential tokens, clearly in favor or against cancellation, were selected and recorded in a Document-Term Matrix. Subsequently, a Word Search was applied, which, without oversampling, yielded results similar to the other two methods used later: Random Forest and Support Vector Machine.

In the subsequent classification of cancellations into individual reasons, four separate binary models were created in a ``one vs. rest" manner. Each model examined whether a cancellation would fit into one of the reasons. In case of a negative result, it was concluded that no reason was provided in the given cancellation.

The methods used remained consistent, and the results were exemplified with the most frequently occurring reason, ``Financial reasons." All three methods produced similar results as in the classification of ``Cancellation" and ``Non-cancellation." However, unlike the initial classification, the models were uniformly less accurate and not as effective as before. In this case, the Word Search emerged as the best method for three out of four reasons. Nevertheless, it can be summarized that a much larger number of letters with reasons is needed, as the current quantity is far from being sufficient to adequately recognize the reasons.

\subsubsection*{Other Modeling Approaches} \label{othermodels}

Another issue, which was not extensively discussed in Section~\ref{tm}, is that incorrectly read words can distort the analysis. One potential approach is to consider the Levenshtein distance\footnote{The Levenshtein distance between two strings is the minimum number of insertions, substitutions, and deletions required to transform one string into another (see \citealp{levenshtein}).}
between each pair of strings. For instance, if this distance is one, the words could be considered identical. However, since there are genuinely different words that have a Levenshtein distance of one but differ only in one letter, this approach is not chosen here. Nevertheless, a more in-depth analysis of this matter could contribute to cleaning the data to some extent.

In addition to this adjustment, other methods could be considered. More recent approaches to classification, particularly in cases of imbalanced data, include ensemble methods such as XGBoost \citep{chen2016xgboost} as well as resampling techniques like SMOTE \citep{SMOTE}. These methods have shown strong performance in comparable contexts and could serve as valuable benchmarks for future work. Furthermore, more sophisticated approaches based on neural networks may provide additional opportunities for improvement.

Due to the fact that all methods and analyses required less than one hour each to be executed on the given dataset, which already comprised a substantial number of documents, there was neither a strong necessity nor sufficient incentive to further optimize the computational implementation. However, this aspect may become increasingly relevant when applying these methods to significantly larger datasets. In such cases, optimizing the programming approach and improving computational efficiency could represent a worthwhile direction for future research.

\subsubsection*{Approach to New Data} 

The long-term goal is to apply these processes to new customer letters. This can be achieved by training the models with current data, or in the case of the Word Search, determining new signal tokens for further classification based on the current training and test data.

Another option is to evaluate new data using existing models. For this purpose, the Word Search is chosen as the final model. In these analyses, it proved to be the best model for identifying cancellation reasons and demonstrated similar accuracy in the classification of cancellations as the other models (see \autoref{kfindvgl} and \autoref{Gfindvgl}). Additional advantages include a significant reduction in text data analysis in further analyses. A new text only needs to be reduced to its word stems before being put into the classification model. Additionally, any numbers occurring in the text need to be removed. However, a drawback of the Word Search is that it does not provide probabilities.

In the classification process, a Word Search is initially used to determine whether a cancellation is present or not. For documents not classified as cancellations, it is noted that they are not cancellations and they are not further processed. For documents classified as cancellations, each of the four reasons is checked to see if it applies. All identified reasons are then output for each business transaction. If no reason or another reason is present, that information is also provided. Since a Word Search was chosen as the final model, cancellation probabilities cannot be provided here, which would be possible with the selection of another model.

\subsubsection*{Sequential and Temporal Patterns} 

A deeper exploration of sequential patterns and temporal data, such as customers writing multiple letters over time, could provide valuable additional insights and potentially improve model performance. As already mentioned in Section~\ref{acquisition}, the present study relied only on letters from the last few years that were specifically requested and made available. In a more comprehensive analysis, incorporating the entire sequence of customer correspondence would be a meaningful extension, as it could reveal behavioral dynamics over time and thus contribute to a more nuanced understanding of customer actions.

\clearpage

\begin{appendix}

\section{Appendix}\label{sec:append}

\subsection{Additions to the Problem Statement and Methods}

\begin{table}[h!]
\centering
\begin{tabular}{lr}
\toprule
Business Transaction & Frequency \\ \midrule
Cancellation & $3,477$ \\
Premium Suspension & $1,512$ \\
Premium Waiver & $641$ \\
Increase & $2,366$ \\
Policyholder Change & $6,579$\\ \bottomrule
Total Number & $14,575$
\end{tabular}
\caption{Overview of the available customer letters and their corresponding triggering reasons based on the original classification}
\label{overview}
\end{table}

\begin{table}[h!]
\centering
\begin{tabular}{lr}
\toprule
Package Name & Version  \\ \midrule
\texttt{collections} & 3.9 \\  
\texttt{datetime} & 4.0.1 \\ 
\texttt{dill}              &                 0.3.4 \\
\texttt{imbalanced-learn}   &                0.9.0\\
\texttt{matplotlib}          &               3.4.3\\
\texttt{numpy}                &              1.20.3\\
\texttt{os} & 0.1.4 \\ 
\texttt{pandas}                &             1.3.4\\
\texttt{pdf2image}              &            1.16.0\\
\texttt{pikepdf}                 &           4.3.0\\
\texttt{pytesseract}              &          0.3.9\\
\texttt{random} & 3.9 \\ 
\texttt{scikit-learn}               &        0.24.2 \\
\texttt{snowballstemmer}             &       2.1.0\\
\texttt{spyder}                       &      5.1.5\\
\texttt{spyder-kernels}                &     2.1.3\\
\texttt{stop-words}                     &    2018.7.23\\
\texttt{warnings} & 3.9 \\ 
\texttt{wordcloud  }                     &   1.8.1\\
\texttt{XlsxWriter}                       &  3.0.1 \\ \bottomrule
\end{tabular}
\caption{Overview of the Packages Used in \texttt{Python}}
\label{Paketübersicht}
\end{table}
\pagebreak

\subsection{Additions to the Classification}

\begin{table}[h!]
\begin{tabular}{lrrr}
\toprule
& \multicolumn{2}{c}{Predicted Class}     &      \\ \cmidrule{2-3}
True Class& Cancellation & No Cancellation & $\sum$      \\ \midrule
Cancellation &  $3,792$  &$0$            & $3,792$\\
No Cancellation & $660$   &   $6,924$   &    $7,584$   \\ 
$\sum$ & $4,452$ & $6,924$ & $11,376$ \\ \bottomrule
\end{tabular}
\centering
\caption[Results of classifying all documents with the large Random Forest model on the training data]{Results of classifying all documents with the large Random Forest model on the training data with sensitivity$=1$ and specificity$=0.913$}
\label{RF2Traintab}
\end{table}

\begin{table}[h!]
\begin{tabular}{lrrr}
\toprule
& \multicolumn{2}{c}{Predicted Class}     &      \\ \cmidrule{2-3}
True Class& Cancellation & No Cancellation & $\sum$      \\ \midrule
Cancellation &  $187$   &    $190$       &$377$ \\
No Cancellation & $1,616$  &   $1,635$   & $3,251$      \\ 
$\sum$ & $1,803$ & $1,825$ & $3,628$ \\ \bottomrule
\end{tabular}
\centering
\caption[Results of classifying all documents with the large Random Forest model on the test data]{Results of classifying all documents with the large Random Forest model on the test data with sensitivity$=0.496$ and specificity$=0.503$ } 
\label{RF2Testtab}
\end{table}

\begin{table}[h!]
\begin{tabular}{lrrr}
\toprule
& \multicolumn{2}{c}{Predicted Class}     &      \\ \cmidrule{2-3}
True Class& Cancellation & No Cancellation & $\sum$      \\ \midrule
Cancellation &  $3,503$		   & $289$           & $3,792$\\
No Cancellation & $206$  & $7,378$     &      $7,584$ \\ 
$\sum$ & $3,709$ & $7,667$ & $11,376$\\ \bottomrule
\end{tabular}
\centering
\caption[Results of classifying all documents with the small Random Forest model on the training data]{Results of classifying all documents with the small Random Forest model on the training data with sensitivity$=0.924$ and specificity$=0.973$} 
\label{RFTraintab}
\end{table}
\begin{table}[h!]

\begin{tabular}{lrrr}
\toprule
& \multicolumn{2}{c}{Predicted Class}     &      \\ \cmidrule{2-3}
True Class& Cancellation & No Cancellation & $\sum$      \\ \midrule
Cancellation &  $344$   &    $33$       & $377$\\
No Cancellation & $77$  &   $3,174$   &  $3,251$     \\ 
$\sum$ & $421$ & $3,207$ & $3,628$ \\ \bottomrule
\end{tabular}
\centering
\caption[Results of classifying all documents with the small Random Forest model on the test data]{Results of classifying all documents with the small Random Forest model on the test data with sensitivity$=0.912$ and specificity$=0.976$} 
\label{RFTesttab}
\end{table}

\begin{table}[h!]
\begin{tabular}{lrrr}
\toprule
& \multicolumn{2}{c}{Predicted Class}     &      \\ \cmidrule{2-3}
True Class& Cancellation & No Cancellation & $\sum$      \\ \midrule
Cancellation &  $3,792$  &$0$            & $3,792$\\
No Cancellation & $660$   &   $6,924$   &    $7,584$   \\ 
$\sum$ & $4,452$ & $6,924$ & $11,376$ \\ \bottomrule
\end{tabular}
\centering
\caption[Results of classifying all documents with the large SVM model on the training data]{Results of classifying all documents with the large SVM model on the training data with sensitivity$=1$ and specificity$=0.913$}
\label{SVM2Traintab}
\end{table}

\begin{table}[h!]
\begin{tabular}{lrrr}
\toprule
& \multicolumn{2}{c}{Predicted Class}     &      \\ \cmidrule{2-3}
True Class& Cancellation & No Cancellation & $\sum$      \\ \midrule
Cancellation &  $54$   &    $323$       & $377$\\
No Cancellation & $444$  &  $2,807$    &   $3,251$    \\ 
$\sum$ & $498$ & $3,130$ & $3,628$ \\ \bottomrule
\end{tabular}
\centering
\caption[Results of classifying all documents with the large SVM model on the test data]{Results of classifying all documents with the large SVM model on the test data with sensitivity$=0.143$ and specificity$=0.863$} 
\label{SVM2Testtab}
\end{table}

\begin{table}[h!]
\begin{tabular}{lrrr}
\toprule
& \multicolumn{2}{c}{Predicted Class}     &      \\ \cmidrule{2-3}
True Class& Cancellation & No Cancellation & $\sum$      \\ \midrule
Cancellation &  $3,444$   &     $348$      & $3,792$ \\
No Cancellation & $239$  &     $7,345$ &  $7,584$     \\ 
$\sum$ & $3,683$ & $7,693$ & $11,376$\\ \bottomrule
\end{tabular}
\centering
\caption[Results of classifying all documents with the small SVM model on the training data]{Results of classifying all documents with the small SVM model on the training data with sensitivity$=0.908$ and specificity$=0.968$} 
\label{SVMTraintab}
\end{table}

\begin{table}[h!]
\begin{tabular}{lrrr}
\toprule
& \multicolumn{2}{c}{Predicted Class}     &      \\ \cmidrule{2-3}
True Class& Cancellation & No Cancellation & $\sum$      \\ \midrule
Cancellation &   $349$  &          $28$ & $377$ \\
No Cancellation & $104$  &    $3,147$  &  $3,251$     \\ 
$\sum$ & $453$ & $3,175$ & $3,628$\\ \bottomrule
\end{tabular}
\centering
\caption[Results of classifying all documents with the small SVM model on the test data]{Results of classifying all documents with the small SVM model on the test data with sensitivity$=0.926$ and specificity$=0.968$} 
\label{SVMTesttab}
\end{table}

\begin{table}[h!]
\begin{tabular}{lrrr}
\toprule
& \multicolumn{2}{c}{Predicted Class}     &      \\ \cmidrule{2-3}
True Class&Financial Reason   &No Financial Reason   & $\sum$      \\ \midrule
Financial Reason&   $13$  &      $5$     & $18$\\
No Financial Reason & $10$  &  $851$    &    $861$   \\ 
$\sum$ &$23$&$856$ & $879$\\ \bottomrule
\end{tabular}
\centering
\caption[Result of the Classification of Cancellations with Word Search on Training Data]{Result of the classification of cancellations with Word Search on training data with sensitivity$=0.722$ and specificity$=0.988$}
\label{FWSTraintab}
\end{table}

\begin{table}[h!]
\begin{tabular}{lrrr}
\toprule
& \multicolumn{2}{c}{Predicted Class}     &      \\ \cmidrule{2-3}
True Class& Financial Reason  & No Financial Reason  & $\sum$      \\ \midrule
Financial Reason&   $2$  &     $6$      & $8$\\
No Financial Reason & $3$  &   $367$   &  $370$     \\ 
$\sum$ &$5$& $373$& $378$\\ \bottomrule
\end{tabular}
\centering
\caption[Result of the Classification of Cancellations with Word Search on Test Data]{Result of the classification of cancellations with Word Search on test data with sensitivity$=0.25$ and specificity$=0.992$}
\label{FWSTesttab}
\end{table}

\begin{table}[h!]
\begin{tabular}{lrrr}
\toprule
& \multicolumn{2}{c}{Predicted Class}     &      \\ \cmidrule{2-3}
True Class& Financial Reason   & Non-Financial Reason   & $\sum$      \\ \midrule
Financial Reason &   $58$  &    $14$       &$72$ \\
Non-Financial Reason &  $44$ &  $937$    &  $981$     \\ 
$\sum$ &$106$& $951$& $1,053$\\ \bottomrule 
\end{tabular}
\centering
\caption[Result of the cancellation classification with Random Forest on training data]{Result of the cancellation classification with Random Forest on training data with sensitivity$=0.806$ and specificity$=0.955$} 
\label{FRFTraintab}
\end{table}

\begin{table}[h!]
\begin{tabular}{lrrr}
\toprule
& \multicolumn{2}{c}{Predicted Class}     &      \\ \cmidrule{2-3}
True Class& Financial Reason   & Non-Financial Reason   & $\sum$      \\ \midrule
Financial Reason&   $2$  &  $6$         &$8$ \\
Non-Financial Reason &  $11$ &  $359$    &  $370$     \\ 
$\sum$ &$13$&$365$ & $378$\\ \bottomrule
\end{tabular}
\centering
\caption[Result of the cancellation classification with Random Forest on test data]{Result of the cancellation classification with Random Forest on test data with sensitivity$=0.25$ and specificity$=0.970$} 
\label{FRFTesttab}
\end{table}

\begin{table}[h!]
\begin{tabular}{lrrr}
\toprule
& \multicolumn{2}{c}{Predicted Class}     &      \\ \cmidrule{2-3}
True Class& Financial Reason  & Non-Financial Reason    & $\sum$      \\ \midrule
Financial Reason&   $52$  &   $20$        & $72$\\
Non-Financial Reason &  $6$ &    $975$  &    $981$   \\ 
$\sum$ &$58$& $995$& $1,053$\\ \bottomrule
\end{tabular}
\centering
\caption[Result of the cancellation classification with SVM on training data]{Result of the cancellation classification with SVM on training data with sensitivity$=0.722$ and specificity$=0.994$} 
\label{FSVMTraintab}
\end{table}

\begin{table}[h!]
\begin{tabular}{lrrr}
\toprule
& \multicolumn{2}{c}{Predicted Class}     &      \\ \cmidrule{2-3}
True Class& Financial Reason  & Non-Financial Reason    & $\sum$      \\ \midrule
Financial Reason&   $1$  &    $7$       & $8$\\
Non-Financial Reason & $0$  &    $370$  &  $370$     \\ 
$\sum$ &$1$& $377$& $378$\\ \bottomrule
\end{tabular}
\centering
\caption[Result of the cancellation classification with SVM on test data]{Result of the cancellation classification with SVM on test data with sensitivity$=0.125$ and specificity$=1$}
\label{FSVMTesttab}
\end{table}


\begin{table}
\begin{subtable}{\textwidth}
\begin{tabular}{lrrr}
\toprule
& \multicolumn{2}{c}{Predicted Class}     &      \\ \cmidrule{2-3}
True Class& Job Change   & No Job Change & $\sum$      \\ \midrule
Job Change&  $1$   &   $6$        & $7$\\
No Job Change&  $16$ &   $355$   &  $371$     \\ 
$\sum$ &$17$& $361$& $378$\\ \bottomrule
\end{tabular}
\centering
\caption{Result of the cancellation classification with Word Search with sensitivity$=0.143$ and specificity$=0.957$}
\label{BWS}
\end{subtable}
\ \\
\\
\begin{subtable}{\textwidth}
\begin{tabular}{lrrr}
\toprule
& \multicolumn{2}{c}{Predicted Class}     &      \\ \cmidrule{2-3}
True Class& Job Change   & No Job Change & $\sum$      \\ \midrule
Job Change &  $0$   &$7$           & $7$\\
No Job Change &  $5$ &    $366$  &  $371$     \\ 
$\sum$ &$5$& $373$& $378$\\ \bottomrule
\end{tabular}
\centering
\caption{Result of the cancellation classification with Random Forest with sensitivity$=0$ and specificity$=0.986$}
\label{BRF}
\end{subtable}
\ \\
\\
\begin{subtable}{\textwidth}
\begin{tabular}{lrrr}
\toprule
& \multicolumn{2}{c}{Predicted Class}     &      \\ \cmidrule{2-3}
True Class& Job Change   & No Job Change & $\sum$      \\ \midrule
Job Change &  $0$   &$7$           & $7$\\
No Job Change &  $5$ &    $366$  &  $371$     \\ 
$\sum$ &$5$& $373$& $378$\\ \bottomrule
\end{tabular} 
\centering
\caption{Result of the cancellation classification with SVM with sensitivity$=0$ and specificity$=0.986$}
\label{BSVM}
\end{subtable}
\caption{Result tables for the cancellation reason Job Change (all classification tables after classifying the test data)}
\label{BALL}
\end{table}


\begin{table}
\begin{subtable}{\textwidth}
\begin{tabular}{lrrr}
\toprule
& \multicolumn{2}{c}{Predicted Class}     &      \\ \cmidrule{2-3}
True Class& Retirement  & No Retirement & $\sum$      \\ \midrule
Retirement &   $2$  &  $3$         & $5$\\
No Retirement & $53$  &  $320$    &  $373$     \\ 
$\sum$ &$55$&$323$ & $378$\\ \bottomrule
\end{tabular}
\centering
\caption{Result of the cancellation classification with Word Search with sensitivity$=0.4$ and specificity$=0.858$}
\label{RWS}
\end{subtable}
\ \\
\\
\begin{subtable}{\textwidth}
\begin{tabular}{lrrr}
\toprule
& \multicolumn{2}{c}{Predicted Class}     &      \\ \cmidrule{2-3}
True Class& Retirement  & No Retirement & $\sum$      \\ \midrule
Retirement &   $2$  &  $3$         & $5$\\
No Retirement & $2$  &  $371$   &  $373$     \\  
$\sum$ &$5$&$374$ & $378$\\ \bottomrule
\end{tabular}
\centering
\caption{Result of the cancellation classification with Random Forest with sensitivity$=0.4$ and specificity$=0.995$} 
\label{RRF}
\end{subtable}
\ \\
\\
\begin{subtable}{\textwidth}
\begin{tabular}{lrrr}
\toprule
& \multicolumn{2}{c}{Predicted Class}     &      \\ \cmidrule{2-3}
True Class& Retirement  & No Retirement & $\sum$      \\ \midrule
Retirement &   $2$  &  $3$         & $5$\\
No Retirement & $0$  &  $373$   &  $373$     \\ 
$\sum$ &$2$&$376$ & $378$\\ \bottomrule
\end{tabular}
\centering
\caption{Result of the cancellation classification with SVM with sensitivity$=0.4$ and specificity$=1$}
\label{RSVM}
\end{subtable}
\caption{Result tables for the cancellation reason Retirement  (all classification tables after classifying the test data)}
\label{RALL}
\end{table}


\begin{table}
\begin{subtable}{\textwidth}
\begin{tabular}{lrrr}
\toprule
& \multicolumn{2}{c}{Predicted Class}     &      \\ \cmidrule{2-3}
True Class& Death   & No Death & $\sum$      \\ \midrule
Death& $3$ & $2$       &$5$ \\
No Death & $1$& $372$  & $373$      \\
$\sum$ &$4$& $374$& $378$\\ \bottomrule
\end{tabular}
\centering
\caption{Result of the cancellation classification with Word Search with sensitivity$=0.6$ and specificity$=0.997$}
\label{TWS}
\end{subtable}
\ \\
\\
\begin{subtable}{\textwidth}
\begin{tabular}{lrrr}
\toprule
& \multicolumn{2}{c}{Predicted Class}     &      \\ \cmidrule{2-3}
True Class& Death   & No Death & $\sum$      \\ \midrule
Death& $3$ & $2$       &$5$ \\
No Death & $1$& $372$  & $373$      \\
$\sum$ &$4$& $374$& $378$\\ \bottomrule
\end{tabular}
\centering
\caption{Result of the cancellation classification with Random Forest with sensitivity$=0.6$ and specificity$=0.997$} 
\label{TRF}
\end{subtable}
\ \\
\\
\begin{subtable}{\textwidth}
\begin{tabular}{lrrr}
\toprule
& \multicolumn{2}{c}{Predicted Class}     &      \\ \cmidrule{2-3}
True Class& Death   & No Death & $\sum$      \\ \midrule
Death& $3$ & $2$       &$5$ \\
No Death & $1$& $372$  & $373$      \\
$\sum$ &$4$& $374$& $378$\\ \bottomrule
\end{tabular}
\centering
\caption{Result of the cancellation classification with SVM with sensitivity$=0.6$ and specificity$=0.997$} 
\label{TSVM}
\end{subtable}
\caption{Result tables for the cancellation reason Death (all classification tables after classifying the test data)}
\label{TALL}
\end{table}
\end{appendix}

\clearpage

\bibliography{schriftstueckepaper.bib}

\end{document}